 \documentclass[oldversion]{aa}

\usepackage{natbib}
\bibpunct{(}{)}{;}{a}{}{,} 
 
\begin{document} 
 
\title{Comment on "PIC simulations of circularly polarised Alfv\'en wave phase mixing: 
A new mechanism for electron acceleration in collisionless plasmas" by Tsiklauri et al.} 
 
\author{F. Mottez \inst{1} \and V. G\'enot \inst{2} \and P. Louarn \inst{2}} 
\institute{Centre d'\'etude des Environnements Terrestre et Plan\'etaires, 
10-12 Av. de l'Europe, 78140 V\'elizy, France;\\ 
Now at LUTH, Observatoire de Paris, 5 place Jules Janssen, 92195 Meudon, France.
\and Centre d'Etude Spatiale des Rayonnements, 9 Av. Colonel Roche, 31400 Toulouse, France.}

\titlerunning{Comment on "PIC simulations of cicrularly..."}
\authorrunning{Mottez et al.}


\date{Received September 20, 2005; accepted November 18, 2005}



\abstract{Tsiklauri et al. recently published a theoretical model of electron acceleration by Alfv\'en waves 
in a nonuniform collisionless plasmas. We compare their work with a series of results published earlier 
by an another team, of which Tsiklauri et al. were probably unaware.
We show that these two series of works, apparently conducted independently, lead 
to the same conclusions. This reinforces the theoretical consistency of the model.}

\maketitle
Two recent papers published by \citet{Tsiklauri_2005NJPh,Tsiklauri_2005AA}, 
based on kinetic numerical simulations, show a mechanism of astrophysical relevance that could explain
the acceleration of electrons in a magnetized collisionless plasma.

The idea is the following : Alfv\'en waves (AW) are often observed or inferred 
in astrophysical plasmas ; it has been theoretically demonstrated that 
they could be the source of plasma acceleration. In parallel propagation 
(i.e. with a wave vector
aligned with the local magnetic field) an AW carries no parallel electric field, 
and therefore cannot accelerate particles
in that direction. However when the wave is structured in the perpendicular
direction, i.e. when $k_\perp \ne 0$ and sufficiently large, the AW 
acquires a significant parallel electric field component. 
Oblique AW can form when an initially parallel AW propagates in regions with
perpendicular Alfven velocity (or density) gradients. Such a situation leads to
a distortion of the wavefronts: small perpendicular 
length scales are formed and finally a parallel electric field component appears.

Although not new as claimed by the authors, this mechanism is, in our opinion, interesting. In this comment, 
the results exposed by \citet{Tsiklauri_2005AA} are briefly discussed and compared with other 
works on the same subject 
published earlier by  \citet{Genot_1999,Mottez_2000,Mottez_2001_d,Mottez_2004_a}. 

G\'enot et al. investigated the acceleration of electrons  in the Earth's aurora. 
Following in-situ exploration by satellites, 
the well-constrained observations make  this region appropriate for the testing of theoretical models. 
G\'enot et al. chose a plasma with a high magnetic field, given by a ratio
of the electron cyclotron to the electron plasma period 
$\omega_{ce}/\omega_{pe} =4$, as in the Earth auroral zone. 
Because plasma cavities are observed
in connexion with electron acceleration and plasma turbulence, the gradients delimit a density depletion.
\citet{Tsiklauri_2005AA}   refer to electron acceleration in solar open coronal structures,
the magnetic field is weaker $\omega_{ce}/\omega_{pe} =1$ and they consider a density bump 
(expected, for instance, at the boundary of a coronal loop). 

Do the two works explore fundamentaly different regimes ?  
\citet{Goertz_1984}, using the Maxwell-Vlasov 
equations, performed a linear study of the oblique AW propagation. He showed
that the value of the parallel electric field depends on  the dimensionless parameter
$r=\beta m_i/m_e$ with respect to one.
($\beta=2\mu_0 p/B^2$ is the ratio of the kinetic plasma pressure to the magnetic pressure, and $m_i/m_e$
is the ion to electron mass ratio.) 
In G\'enot et al. $m_i/m_e=100$, $\beta \sim 800^{-1}$ and $r=0.125$. In Tsiklauri
et al. $m_i/m_e=16$, $\beta \sim 50^{-1}$ and $r=0.32$. All the publications discussed in this comment 
deal with the same regime, $r <1$,  the 
inertial AW regime. 

\citet{Genot_1999} made a linear study of the wave propagation in a 
transverse gradient (for $r <1$), and showed that to create a parallel electric field, the characteristic
size of the gradient must be of the order of the electron inertial length $c/\omega_{pe}$. 
The simulations carried out by Tsiklauri et al. and by G\'enot et al. are all in this regime; this allows for almost direct 
comparisons.

\citet{Tsiklauri_2005NJPh} compare their results 
to those of \citet{Hasegawa_1975,Hasegawa_1976}. Many conclusions seem to agree, and
we expect that the differences (appart from non homogeneity, not treated by Hasegawa and Chen) are due to the parameter
$r$ that is  greater than one in \citet{Hasegawa_1975,Hasegawa_1976}. When $r>1$, the characteristic transverse
length is not $c/\omega_{pe}$ but the ion Larmor radius. 

The PIC simulation codes used by G\'enot et al. and Tsiklauri et al. are slightly different.
In \citet{Tsiklauri_2005NJPh,Tsiklauri_2005AA}, the code describes the full dynamics of the electrons, and requires
a small time step (a run requires 8 days). 
In \citet{Mottez_2000,Mottez_2001_b,Mottez_2001_d,Mottez_2004_a}, the code describes the guiding centre dynamics
of the electrons, and if $\omega_{ce}/\omega_{pe} >1$, it allows for larger time steps. 
A simulation typically lasts 10 hours.

The initial conditions are very similar  in \citet{Tsiklauri_2005AA} and \citet{Genot_1999} : a local perturbation 
is propagated by parallel Alfv\'en waves. 
Tsiklauri et al. show that the AW propagation in their kinetic
simulation is like in MHD. In the other works conducted by  
\citet{Mottez_2000,Mottez_2001_b,Mottez_2004_a}, the simulations
are initialised with bi-fluid (therefore almost MHD) Alfv\'en waves filling the whole simulation domain. 
The kink in the ion velocities, observed by Tsiklauri et al. is set initially by G\'enot et al.
Such an initialization
allows one to control the direction of propagation of the wave. 
\citet{Mottez_2000} could show that 
the electron acceleration occurs most efficiently in the direction of propagation of the AW; the bulk velocity 
is translated in the direction of propagation of the wave. But some electrons are (much less) accelerated  
in the reverse direction. As the Alfv\'en velocity is, in our regime, much faster than the electron
thermal velocity, we consider that the acceleration process is non-resonnant (and therefore not a Landau
process), even if second-order resonance effects give more efficient acceleration in the direction of the wave
\citep{Mottez_2001_b}. 

The parallel electric field \citep{Mottez_2000,Mottez_2001_b,Mottez_2004_a} 
shows a spiky structure superimposed on a smooth and less intense electric field. 
The smooth electric field is created in the linear phase of the AW dissipation process
\citep{Genot_1999}, as can be seen in simulations \citep{Mottez_2001_d}. As this 
field exists over large distances along the magnetic field, it has a high potential for accelerating electrons.
Then, when the acceleration begins, many strong and localized quasi-electrostatic high amplitude spikes develop. 
These spikes are probably
the structures seen on the $E_x$ displayed in \citet{Tsiklauri_2005AA}. Their nature and origin 
have been extensively studied \citep{Mottez_2001_d,Mottez_2004_a}. 
They are the consequence of the nonlinear interaction of the 
beams of accelerated electrons and current densities with the bulk of the plasma
(beam-plasma and Buneman instabilities). These spikes have the same 
characteristics as those 
observed near the plasma cavities by auroral satellites. If this process is indeed correctly understood, 
they are a consequence,
and not the cause of the electron acceleration
\citep{Mottez_2004_a}. 

As the magnetic field is lower in the simulations conducted by Tsiklauri et al., the Alfv\'en wave velocity 
is smaller, closer to the electron thermal velocity, and it is possible that resonnant effects 
are more important in the wave particle interaction.

In spite of some minor differences of regime and minor divergences 
in the interpretation of the simulation results, 
the rediscovery of the acceleration process initially proposed by G\'enot et al.
and the similarity of the most important results strenghtens the theoretical relevance of this model.


\bibliographystyle{aa}        
\bibliography{AA_2005_4229.bbl}       


\end{document}